\documentclass[journal]{IEEEtran}
\usepackage{graphicx}
\usepackage{epstopdf} 
\usepackage{verbatim} 
\usepackage{multirow} 
\usepackage{multicol}
\usepackage{amsfonts}
\usepackage{bbm}
\usepackage{subfigure}
\usepackage{colortbl} 
\usepackage{indentfirst}
\usepackage{dcolumn,booktabs}
\usepackage{algorithm}
\usepackage[noend]{algpseudocode}
\usepackage{algorithmicx,algorithm}

\usepackage{algpseudocode}
\usepackage{graphics}
\usepackage{epsfig,color}
\usepackage{amsmath}
\usepackage{amssymb}
\ifCLASSINFOpdf

\else
\fi
\newtheorem{Theorem}{Theorem}
\newtheorem{Corollary}{Corollary}

\newtheorem{Lemma}{Lemma}

\begin{document}

\title{A Fine-Grained Analysis of mmWave Heterogeneous Networks}

\author{
\IEEEauthorblockN{Le Yang, Fu-Chun Zheng and Shi Jin}
}
\maketitle

\begin{abstract}
A fine-grained analysis of the cache-enabled networks is crucial for system design. In this paper, we focus on the meta distribution of the signal-to-interference-plus-noise-ratio (SINR) in the mmWave heterogeneous networks where the base stations (BS) in each tier are modeled as Poisson point process (PPP). By utilizing stochastic geometry, we derive the moments of the conditional success probability, based on which the exact expression of meta distribution and its beta approximation are derived. In addition, key performance metrics, the success probability, the variance of the conditional success probability, the mean local delay and the network jitter are achieved. The distinguishing characteristics of the mmWave communications, including different path loss laws for line-of-sight and non-line-of-sight links and directional beamforming are incorporated into the analysis. The simulation results reveal the impact of the key network parameters, such as blockage parameter, bias factor, number of antenna elements and density on the performance.
\end{abstract}


%
\IEEEpeerreviewmaketitle

\section{Introduction}
\subsection{Motivation}
Accompanying with the rapid proliferation of the communication and electronic technologies, various new applications have emerged, such as autonomous vehicles, virtual reality and augmented reality. As a result, the fifth generation (5G) and beyond 5G are anticipated to provide the massive connectivity for the enormous amount of services \cite{massive-connectivity}. Specifically, the data rate demand of the virtual and augmented realities reaches the order of gigabit-per-second \cite{AR-VR}. In order to meet the increasing data rate demand, millimeter wave (mmWave) communications stands out as a promising approaches to realize the gigabit data rate \cite{mmwave-technology}.

The mmWave spectrum corresponding from 30GHz to 300GHz has attracted considerable attention from both academia and industry due to the huge bandwidth. However, the mmWave communications suffer from several drawbacks, such as the high path loss caused by its susceptibility to the atmospheric absorption, diffraction and deflection and the poor penetration to the blockages \cite{mmwave-nature}. Fortunately, these limitations could be overcome by deploying directional antennas. Channel measurements have shown that a transmission range of 150-200 meters can be achieved. In addition, much higher data rate and comparable coverage rather than the sub-6GHz networks can be obtained. Another important feature of the mmWave networks is the heterogeneity.

In general, a heterogeneous network consists of $K$ network tiers distinguished by the spatial densities, the transmit powers and the blockage models \cite{blockage-model-1}, \cite{blockage-model-2}. For example, the low-power and high-density small-cell base stations (BSs) coexist with the high-power and low-density large-cell BSs. The small-cell BSs can offload some percentage of users from the congested large-cell BSs to improve the quality of service \cite{QoS}. In order to alleviate the burden of the large-cell BSs, cell range expansion (CRE) technology can be utilized \cite{load-balancing-1}, \cite{load-balancing-2}. In order to evaluate the performance of the mmWave heterogeneous networks, stochastic geometry is utilized as a powerful analyzing tool due to its capability of capturing the variability and irregularity of the node locations in real networks. However, the current analysis mainly focus on the performance of the typical user by spatial averaging. A important performance metric is the success probability, which is defined as the complementary cumulative density function (CCDF) of the signal-to-interference-ratio (SIR). The success probability is obtained by calculating the expectation of the conditional success probability over the underlying point processes modeling the locations of the BSs in all tiers. Therefore, it cannot reflect the performance variation for each individual link. Note that the real deployment of the mmWave heterogeneous networks concerned by the network operators are the questions such as ``What fraction of users can achieve the STP of at least $x$ (an arbitrary percentage value)?''. Unfortunately, the success probability only answers the question of ``On average what fraction of users experience successful transmission?''. To overcome this drawback and obtained a fine-grained analysis on the SIR distribution, the meta distribution is introduced \cite{md-maenggi}, which is defined as the CCDF of the success probability.

\subsection{Related Works}
The mmWave networks has been extensively investigated in the previous literatures by utilizing tools from stochastic geometry. The authors in \cite{Bai} proposed a analytical framework to analyze the mmWave cellular networks. The expression of coverage probability and the achievable rate were derived and the simplified expression were obtained by utilizing the equivalent LOS ball model in the dense BS deployment scenario. In \cite{impact}, the authors investigated the impact of the directional antenna arrays by proposing two antenna patterns with analytical tractability and desired accuracy to approximate the actual antenna pattern and derived the corresponding coverage probability in the mmWave ad hoc and cellular networks. In \cite{mmwave-mac}, two types of heterogeneity, i.e., spectrum heterogeneity and deployment heterogeneity, were considered in the mmWave networks. In spectrum heterogeneity, the users used the low frequencies for control message exchange while the higher frequencies are used for the data communication. For deployment heterogeneity, two deployment scenarios, i.e., the stand-alone scenario and the integrated scenario, were introduced. All tiers operated in the mmWave frequency bands in the stand-alone scenario while the sub-6GHz networks coexisted with the mmWave networks in the integrated scenario.

The coverage probability of the stand-alone mmWave heterogeneous networks are studied by a variety of research works. In \cite{K-tier-Renzo}, the authors proposed a generalized mathematical framework for the analysis of the mmWave heterogeneous networks where the coverage probability and the average rate were obtained. In \cite{Bai} and \cite{K-tier-Renzo}, a distance-dependent line-of-sight (LOS) probability function is utilized where the LOS and non-line-of-sight (NLOS) BSs are distributed as two independent Poisson point process (PPP). The LOS/NLOS state of each link is distinguished by the path loss model and the small-scale fading where the Nakagami fading with different parameters are assumed for LOS and NLOS links. In \cite{K-tier}, the authors obtained the expression of the coverage probability and energy efficiency in the mmWave heterogeneous networks by utilizing multi-ball approximation for the blockage model. In \cite{low-complexity}, the authors proposed a low complexity BS selection scheme consisting of two-level procedures and provided the analytical and asymptotic expressions of the coverage probability corresponding to three BS pre-selection policies in the mmWave heterogeneous networks. In \cite{cooperative},  the benefit of the BS cooperation in the downlink heterogeneous mmWave cellular networks is investigated.

The coverage or success probability of the coexisting sub-6GHz and mmWave networks has been analyzed in the previous literatures. In
\cite{tractable}, the uplink-downlink signal-noise-ratio (SNR) and rate distribution of the self-backhaul mmWave cellular networks were theoretically elaborated. The authors in \cite{downlink} characterized the uplink and downlink cell association strategies and demonstrated the superiority of the proposed decoupled cell association strategies over the traditional coupled approach in the hybrid networks where the sub-6GHz macro BSs coexisted with the mmWave small-cell BSs. Note that \cite{tractable} and \cite{downlink} utilized the Rayleigh fading as the small-scale fading for the analytical tractability.

The above works only analyze the coverage or success probability of the mmWave networks without delving into the performance variation for each link. To obtain a fine-grained analysis on the network performance, the meta distribution was proposed in \cite{md-maenggi}. The meta distribution of the cellular networks has been investigated extensively in previous works. In \cite{md-maenggi}, where the moments of the conditional success probability, the exact expression and approximation of the meta distribution for the cellular networks and bipolar networks were derived, respectively. The exact analytical expressions and the beta approximations of the meta distribution have since been obtained in various other scenarios, including the heterogeneous networks \cite{md-hetnet},  non-Poisson networks \cite{md-cellular-networks}, D2D communications \cite{md-D2D}, coordinated multipoint transmission \cite{md-bs-cooperation}, non-orthogonal multi-access \cite{md-NOMA} and fractional power control \cite{md-power-control}. In \cite{Deng-fine-grained}, the meta distribution and beta approximation of the SIR in the mmWave D2D networks was derived.

\subsection{Contributions}
Although the meta distribution of SIR is investigated in various previous works, the meta distribution in mmWave heterogeneous networks still remains to be studied. In this paper, we develop a meta distribution analytical framework for the mmWave heterogeneous networks. Unlike \cite{Deng-fine-grained} where the simplified Rayleigh fading channel is assumed and \cite{md-hybrid} where the equivalent ball model is utilized to approximate the LOS probability for each link, we utilize the LOS probability function to approximate the LOS probability of each link and Nakagami fading with different parameters are assumed for the small-scale fading of LOS/NLOS links. Note that the Rayleigh fading is a special case of the Nakagami fading when the shape parameter equals to 1.

Our contribution are summarized as follows
\begin{enumerate}
\item Different from \cite{K-tier-Renzo}, \cite{Bai} and \cite{K-tier} where only the approximation of the success probability in the mmWave heterogeneous networks are derived, we characterize the exact expression of the success probability for each tier. In addition, the simplified expression of the success probability under the special case where the blockage parameters are sufficiently small is derived.
\item The moments and the variance of the conditional success probability are derived when the user is associated with a LOS/NLOS BS in each tier. Based on the results from the moments of the conditional success probability, the exact expression and the beta approximation of the meta distribution of the SINR and the date rate are obtained. In addition, the mean local delay and the variance of the local delay (network jitter) are derived.
\item The numerical results reveal the effect of the directional antenna array gain on the variance of the success probability. The 
\end{enumerate}

\section{System Model}
We consider a downlink scenario in the mmWave heterogeneous networks consisting of $K$ tiers. Denote $\mathcal{K}\triangleq\{1,2,\cdots,K\}$. The BSs in each tier $k\in\mathcal{K}$ are assumed to be distributed as a homogeneous Poisson point process (PPP) $\Phi_k$ with density $\lambda_k$. The BSs in the $k$th tier transmit with power $P_k$. We assume that the BSs in all tiers operate over the same mmWave frequency band and the bandwidth is $W$. Without loss of generality, we study the performance of the typical user $u_0$ located at the origin according to Slivnyak's theorem \cite{slivnyak}.

The wireless channel in the mmWave network is characterized by the large-scale and small-scale fading. We first analyze the large-scale fading. Since the signal is sensitive to the blockages in the mmWave environment, the BSs can be line-of-sight (LOS) or non-line-of-sight (NLOS) based on that whether there is blockage intersecting the link between $u_0$ and the BS. If there is no blockage intersecting the link between the link between $u_0$ and the BS, the corresponding link is LOS. Otherwise, the link is NLOS. The probability that a link between $u_0$ and a BS located at distance $x$ is considered to be LOS is defined as the LOS probability function $p_L{x}=e^{-\beta_kx}$, where $\beta_k$ is the blockage parameter for the $k$th tier and determined by the average size and density of the blockages. In addition, the probability that a link with length $x$ is NLOS is given by $p_N(x)=1-p_L(x)$. The received signal is attenuated due to the path loss and the path loss exponent is utilized to measure the severity of the attenuation. To distinguish the LOS and NLOS states for an arbitrary link, different path loss laws is employed for the LOS and NLOS links as follows
\begin{equation}
L(x)=
\begin{cases}
\kappa_Lx^{\alpha_L}\ \ \ \text{with probability}\ p_L(x)\\
\kappa_Nx^{\alpha_N}\ \ \ \text{with probability}\ p_N(x)
\end{cases}
\end{equation}
where $\kappa_L$ and $\kappa_N$ are the intercepts for the LOS and NLOS link at 1 meter, $\alpha_L$ and $\alpha_N$ the path loss exponent for the LOS and NLOS link.

For the small-scale fading, we denote by $h_{k,i}$ the small-scale fading term of link $i$ in the $k$th tier and assume Nakagami fading with the probability density function (PDF) $f(h)=\frac{2M_{\rho}^{M_{\rho}}h^{2M_{\rho}-1}}{\Gamma(M_{\rho})}e^{-M_{\rho}h^2}$ for each link. Specifically, the Nakagami fading with parameter $M_L$ and $M_N$ is applied to the LOS and NLOS links, respectively. Therefore, $|h_{k,i}|^2$ is a Gamma distributed random variable. Note that the small-scale fading in mmWave networks is less severe than the sub-6GHz networks due to the directional antenna array and a large $M_L$ can be utilized to approximate the fading with small variance for the LOS link. When $M_{\rho}=1,\rho\in\{L,N\}$, the small-scale fading reduces to the Rayleigh fading. With the deployment of the antenna array, the BSs perform beamforming where the main lode directs towards the dominant propagation path and the side lodes direct towards other directions. In addition, for analytical tractability, we utilize a sector model to approximate the array antenna pattern \cite{sector-1}-\cite{sector-3}. The beam direction of the interfering BSs is uniformly distributed on $[0,2\pi]$. The antenna array gain between $u_0$ and an interfering BS is given by
\begin{equation}
G=
\begin{cases}
\Psi, &\text{with probability}\quad \frac{\delta}{2\pi},\\
\psi, &\text{with probability}\quad \frac{1-\delta}{2\pi},
\end{cases}
\end{equation}
where $\Psi$ denotes the main lobe gain, $\psi$ the side lobe gain, $\delta$ the half power beamwidth. We assume that the angle of arrival (AoA) can be estimated at the BS and the antenna orientation steering can be adjusted by utilizing the estimated AoA. Therefore, a perfect alignment is assumed between $u_0$ and its serving BS and the maximum array gain $\Psi$ can be achieved for the link between $u_0$ and its serving BS.

$u_0$ is associated with the BSs providing the strongest signal power among all tiers. Note that the user association is jointly affected by the transmit power of the BS and the path loss of link between $u_0$ and the BS. Due to the LOS/NLOS state of the links between $u_0$ and the BSs, the nearest BS is not necessarily the BSs providing the strongest power in the $k$th tier. In addition, CRE is adopted to offload some user from the large-cell BSs to the small-cell BSs in order to alleviate the burden of the large-cell BSs and enhance the overall performance of the heterogeneous networks \cite{load-balancing-1}, \cite{load-balancing-2}. Therefore, $u_0$ is associated with a BS based on the maximum biased received signal strength, which can be mathematically described as follows
\begin{equation}
P_kB_kG_kL_k(x)^{-1}>P_jB_jG_jL_{j,\min}(x)^{-1}
\end{equation}
where $L_{j,\min}(x)$ denotes the minimum path loss between $u_0$ and the BSs in the $j$th tier.

The SINR is given by
\begin{equation}
\text{SINR}=\frac{P_kh_{k,0}L_k(x)^{-1}}{\sigma^2+\sum_{j=1}^{K}\sum_{i\in\Phi_j\backslash B_{k,0}}P_jG_{j,i}h_{j,i}L_{j,i}^{-1}(x)}
\end{equation}
where $G_{j,i}$, $h_{j,i}$ and $L_{j,i}$ denote array gain, small-scale fading and the path loss for the interfering link, the $\sigma^2$ the thermal noise.
The meta distribution of the SINR $\bar{F}_{\mathcal{P}}(x)$ is defined as the complementary cumulative distribution function (CCDF) of the conditional success probability
\begin{equation}
\mathcal{P}(\theta)=\mathbbm{P}(\text{SIR}>\theta|\Phi),
\end{equation}
which is the CCDF of the SIR for $u_0$ conditioned on the realization of $\Phi$. Henceforth, the meta distribution is given by \cite{md-maenggi}
\begin{equation}
\bar{F}_{\mathcal{P}}(y)\triangleq\mathbbm{P}(\mathcal{P}(\theta)>y),\ y\in[0,1].
\end{equation}

Due to the ergodicity of the point processes, the meta distribution can be regarded as the fraction of active links with the conditional success probability greater than $x$.

The expression of the meta distribution includes multi-order moments of the conditional success probability. Denoting the $b$-th moment of $\mathcal{P}$ by $M_b$, we can derive the expression of meta distribution by utilizing the Gil-Pelaez theorem. The exact form of the meta distribution is complex and considerable time will be consumed to obtain the final results. In order to facilitate the analysis, the beta distribution is utilized to approximate the meta distribution by matching the first and second moments.

We also analyze the per-link delay consisting of two parts, i.e., the transmission delay and the queuing delay. the retransmission delay is the main component of the transmission delay , which is defined as the number of retransmissions needed until a successful transmission occurs \cite{local-delay}. It can also be called the local delay. We denoted the local delay by $L$ and thus the mean local delay can be written as
\begin{equation}\label{delay-definition}
\mathbbm{E}[L]\overset{(a)}{=}\mathbbm{E}\left[\frac{1}{\mathcal{P}(\theta)}\right]=M_{-1}.
\end{equation}
where (a) follows from the fact that $L$ is geometrically distributed with parameter $\mathcal{P}(\theta)$ conditioned on $\Phi$. Denoting $L|\Phi$ by $L_{\Phi}$, we have
\begin{equation}
\mathbbm{P}(L_{\Phi}=k)=(1-\mathcal{P})^{k-1}\mathcal{P},\ k\in\mathbbm{N}.
\end{equation}
where $\mathcal{P}$ is the conditional success probability. As shown in (\ref{delay-definition}), the mean local delay can be derived by computing the $-1$-st moment of the conditional success probability.

\section{Auxiliary Results}
In this section, we first provide the characteristics of the path loss, i.e., the PDF and the complementary cumulative density function (CCDF) of the path loss, then provide the expression of the association probability. Let $\mathcal{N}_k$ denote the point process of the path loss between $u_0$ and the BSs in the $k$th tier. Therefore, the CCDF and PDF of the path loss can be provided in the following lemma.
\begin{Lemma}\label{lemma-intensity}
The CCDF of the path loss between $u_0$ and its serving BS is given by
\begin{equation}
\bar{F}_{L_k}(x)=\exp(-\Lambda([0,x))),k\in\mathcal{K}
\end{equation}
\end{Lemma}
where $\Lambda_{k}([0,x))$ is defined as in (\ref{Lambda}) on the top of the next page.

\begin{equation}\label{Lambda}
\begin{split}
&\Lambda_k([0,x))=\pi\lambda_k(x/\kappa_N)^{2/\alpha_N}\\
&+2\pi\lambda_k\beta_k^{-2}\left(1-e^{-\beta_k(x/\kappa_L)^{1/\alpha_L}}
\left(1+\beta_k(x/\kappa_L)^{1/\alpha_L}\right)\right)\\
&-2\pi\lambda_k\beta_k^{-2}\left(1-e^{-\beta_k(x/\kappa_N)^{1/\alpha_N}}
\left(1+\beta_k(x/\kappa_N)^{1/\alpha_N}\right)\right).
\end{split}
\end{equation}

\emph{Proof:} See Appendix A.

Unlike the traditional sub-6GHz networks, the blockage parameter $\beta_k$ plays an important role in determining the path loss between $u_0$ and its serving BS. The signal from the serving LOS BS attenuates faster than the NLOS BS. Next, we characterize the path loss between $u_0$ and its serving LOS/NLOS BSs.

\begin{Lemma}
The CCDF of the path loss between $u_0$ and its serving LOS/NLOS BS in the $k$th tier is given by
\begin{equation}
\bar{F}_{L_{k,\rho}}(x)=\exp(-\Lambda_{k,\rho}([0,x))), k\in\mathcal{K},
\end{equation}
where $\rho\in\{\text{LOS},\text{NLOS}\}$ and $\Lambda_{k,\rho}([0,x))$ corresponds to LOS and NLOS, respectively, as shown in (\ref{LOS-intensity}) and (\ref{NLOS-intensity}).
\begin{equation}\label{LOS-intensity}
\begin{split}
&\Lambda_{k,L}([0,x))\\
&=2\pi\lambda_k\beta_k^{-2}\left(1-e^{-\beta_k(x/\kappa_L)^{1/\alpha_L}}
\left(1+\beta_k(x/\kappa_L)^{1/\alpha_L}\right)\right)
\end{split}
\end{equation}

\begin{equation}\label{NLOS-intensity}
\begin{split}
&\Lambda_{k,N}([0,x))=\pi\lambda_k(x/\kappa_N)^{2/\alpha_N}\\
&-2\pi\lambda_k\beta_k^{-2}\left(1-e^{-\beta_k(x/\kappa_N)^{1/\alpha_N}}
\left(1+\beta_k(x/\kappa_N)^{1/\alpha_N}\right)\right)
\end{split}
\end{equation}
\end{Lemma}
\emph{Proof:} $\Lambda_{k,\text{LOS}}([0,x))$ and $\Lambda_{k,\text{NLOS}}([0,x))$ can be derived following the similar steps with the derivation of $\Lambda_{k}([0,x))$.

The PDF of the distance between $u_0$ and its serving BS is given by
\begin{equation}\label{path-loss-pdf}
f_{L_{k,\rho}}=-\frac{d\bar{F}_{L_{k,\rho}}(x)}{dx}=\Lambda_{k,\rho}'([0,x))\exp(-\Lambda_{k,\rho}([0,x)))
\end{equation}

By utilizing the derived results, we can obtain the expression of the association probability in the following lemma.
\begin{Lemma}\label{lemma-association}
The probability that $u_0$ is associated with the $k$th tier is given by
\begin{equation}
A_{k,\rho}=\int_{0}^{\infty}\Lambda_{k,\rho}([0,l_k))e^{\sum_{j=1}^{K}\Lambda_{j}\left([0,\frac{P_jB_JG_j}{P_kB_kG_k}l_k)\right)}\text{d}l_k
\end{equation}
\end{Lemma}
\emph{Proof:} See Appendix B.

From Lemma \ref{lemma-association}, we observe that the association probability is dependent on three sets of parameters, i.e., the physical layer parameters, the antenna array parameters and the mmWave environment parameters. The physical layer parameters include the transmit power $P$ and the BS density $\lambda_k$. The antenna array parameters consist of the main lobe gain $\Psi$, the side lobe gain $\psi$ and the half power beamwidth $\delta$. The mmWave environment parameters consist of the blockage parameters $\beta_k$ and the path loss exponent $\alpha_{\rho},\rho\in\{L,N\}$.

Note that the expression of the association probability is in a complex form. In order to obtain more insight on the effect of the network parameters on the probability that $u_0$ is associated with the $k$th tier, we simplify the expression by utilizing the step function to model the LOS probability of the links between $u_0$ and the BSs in the $k$th tier. In the step function model, we assume that $R_k$ is the maximum length of an LOS link in the $k$th tier. In addition, the LOS probability is 1 for the BSs with the distance $(0,R_k]$ and 0 for the BSs with the distance $(R_k,\infty)$.
\begin{Corollary}
In a two-tier network with the step function to model the LOS probability, the expression of the probability for $u_0$ to be associated with a LOS BS in the $k$th tier is given in (\ref{association-1}) and (\ref{association-2}), as shown at the top of the next page.
\begin{figure*}
\begin{equation}\label{association-1}
A_{1,L}=
\begin{cases}
\frac{\lambda_1P_1G_1B_1}{\sum_{j=1}^{2}\lambda_jP_jG_jB_j}\left(1-e^{\frac{\pi R_1^2}{P_1G_1B_1}\left(\sum_{j=1}^{2}\lambda_jP_jG_jB_j\right)}\right),\ \ \text{if} \sqrt{\frac{P_1G_1B_1}{P_2G_2B_2}}R_{2}>R_1\\
\frac{\lambda_1P_1G_1B_1}{\sum_{j=1}^{2}\lambda_jP_jG_jB_j}\left(1-e^{\frac{\pi R_2^2}{P_2G_2B_2}\left(\sum_{j=1}^{2}\lambda_jP_jG_jB_j\right)}\right)+e^{-\frac{\pi R_2^2}{P_2G_2B_2}\sum_{j=1}^{2}\lambda_jP_jG_jB_j}
-e^{\pi\sum_{j=1}^{2}\lambda_jR_j^2}, \ \ \text{otherwise}
\end{cases}
\end{equation}
\end{figure*}
\begin{figure*}
\begin{equation}\label{association-2}
A_{2,L}=
\begin{cases}
\frac{\lambda_2P_2G_2B_2}{\sum_{j=1}^{2}\lambda_jP_jG_jB_j}\left(1-e^{\frac{\pi R_2^2}{P_2G_2B_2}\left(\sum_{j=1}^{2}\lambda_jP_jG_jB_j\right)}\right),\ \ \text{if} \sqrt{\frac{P_2G_2B_2}{P_1G_1B_1}}R_{1}>R_2\\
\frac{\lambda_1P_1G_1B_1}{\sum_{j=1}^{2}\lambda_jP_jG_jB_j}\left(1-e^{\frac{\pi R_1^2}{P_2G_2B_2}\left(\sum_{j=1}^{2}\lambda_jP_jG_jB_j\right)}\right)+e^{-\frac{\pi R_1^2}{P_2G_2B_2}\sum_{j=1}^{2}\lambda_jP_jG_jB_j}
-e^{\pi\sum_{j=1}^{2}\lambda_jR_j^2}, \ \ \text{otherwise}
\end{cases}
\end{equation}
\end{figure*}
\end{Corollary}

\section{Analysis of Meta Distribution}
In this section, we first provide the exact expression of the success probability, followed by the analytical expression of the moments of the conditional success probability, then the expression of the meta distribution of the SINR distribution is provided. Note that the approximation of the moments of the conditional success probability is also derived.

\subsection{Moments of Conditional Success Probability}
\begin{Theorem}\label{theorem-sp}
Given that $u_0$ is associated with a LOS/NLOS BS in the $k$th tier, the success probability is given by
\begin{equation}\label{equation-sp}
\mathcal{P}_{b,k,\rho}=\frac{1}{A_{k,\rho}}\int_{0}^{\infty}f_{L_{k,\rho}}(l_{k,\rho})\left\|\exp\left(\mathbf{Q}_{M_{\rho}}\right)\right\|_1\text{d}l_{k,\rho}
\end{equation}
where $\exp(\mathbf{A})$ is the matrix exponential, i.e., $\exp(\mathbf{A})=\sum_{k=0}^{\infty}\frac{\mathbf{A}^k}{k!}$, $\|\cdot\|$ is the $l_1$-induced norm, $\mathbf{Q}_{M_{\rho}}$ is a $M_{\rho}\times M_{\rho}$ Toeplitz matrix
\begin{equation}
\mathbf{Q}_{M_L}=
\begin{bmatrix}
q_{k,0}\\
q_{k,1}&q_{k,0}\\
\vdots&\vdots&\ddots\\
q_{M_{\rho}-1}&q_{M_{\rho}-2}&\cdots&q_{k,0},
\end{bmatrix},
\end{equation}
where $q_{k,0}$ and $q_{k,n},n\in\{1,2,\cdots,M_{\rho}-1\}$ is given in (\ref{q-0}) and (\ref{q-m}), as shown at the top of the next page. $(a)_n=a(a+1)\cdots(a+n-1)$ is the rising Pochhammer symbol \cite{integral-table}.
\begin{equation}\label{q-0}
\begin{split}
&q_{k,0}=\sum_{j=1}^{K}\sum_{\nu\in\{L,N\}}\sum_{G\in\{\Psi,\psi\}}\\
&\int_{\frac{P_jB_j}{P_kB_k}l_{k,\rho}}^{\infty}
\left(1-\left(1+\frac{\theta L_{k,\rho}M_{\rho}P_jG}{xP_kG_0M_{\nu}}\right)^{-M_{\nu}}\right)\Lambda_{j,\nu}(\text{d}x)
\end{split}
\end{equation}
\begin{equation}\label{q-m}
\begin{split}
&q_{k,n}=-\sigma^2[2-n]^{+}-\sum_{j=1}^{K}\sum_{\nu\in\{L,N\}}\sum_{G\in\{\Psi,\psi\}}\\
&\frac{(M_{\nu}-1)_nP_jG\theta M_{\rho}L_{k,\rho}}{n!P_kG_0}\int_{0}^{\frac{B_kM_{\rho}}{B_jG_0}}\frac{x^{n-2}}{(1+x)^{M_{\nu}+n}}\Lambda_{j,\nu}(\text{d}x)\\
\end{split}
\end{equation}
\end{Theorem}
\emph{Proof:} See Appendix C.

From Theorem \ref{theorem-sp}, we can observe that the expression of the success probability is in a complex form. The main challenge in the analysis of the multi-antenna networks lies in tackling the high-order derivatives of the Laplace transform. The method in \cite{recursive-1}, \cite{recursive-2} is adopted to obtain the success probability as the $\ell$-induced norm of a Toeplitz matrix representation. Compared with the Fa\`a di Bruno's formula including a large number of products and summations in \cite{Faa-di-Bruno}, the results in (\ref{equation-sp}) is in a more compact form. Different from the heterogeneous networks consisting of BSs equipped with omnidirectional antennas \cite{omnidirectional}, the success probability is dependent on not only the physical layer parameters of all tiers, i.e., the transmit power $P_k$, the BS density $\lambda_k$, but also the antenna array parameters. Note that the parameters related to the mmWave environment play an important role in determining the success probability.

Since both the impact of the LOS and NLOS BSs are considered, the expression in (\ref{equation-sp}) is in a complex form including an integral. In order to facilitate the analysis, we provide the simplified expression of the success probability under the special case where the blockage parameter is sufficiently small.
\begin{Corollary}\label{corollary-sp-los}
In the interference-limited mmWave heterogeneous networks where the blockage parameter for all tier is sufficiently small, i.e., $\beta_k\rightarrow 0,k\in\mathcal{K}$, given $u_0$ is associated with a LOS/NLOS BS, the success probability can be expressed as follows
\begin{equation}
\mathcal{P}_{k,\rho}=V\left\|\left(\sum_{j=1}^{K}T\mathbf{I}-\mathbf{G}_{M_L}\right)^{-1}\right\|_1
\end{equation}
where $V=\sum_{j=1}^{N}\lambda_j\left(\frac{P_jB_j}{P_kB_k}\right)^{\frac{2}{\alpha_L}}$, $T$ is given by
\begin{equation}\label{T}
T=\lambda_j\left(\frac{P_jB_j}{P_kB_k}\right)^{\frac{2}{\alpha_L}}{}_2F_1\left(\frac{2}{\alpha_L},M_L;1-\frac{2}{\alpha_L};-\frac{\theta M_{k,L}B_k}{G_0B_j}\right)
\end{equation}
where $\mathbf{I}$ is a $M_L\times M_L$ identity matrix, $G_{M_L}$ is a $M_L\times M_L$ Toeplitz matrix, which can be expressed as
\begin{equation}
G_{M_L}=
\begin{bmatrix}
g_{k,0}\\
g_{k,1}&g_{k,0}\\
\vdots&\vdots&\ddots\\
g_{M_L-1}&g_{M_{\rho}-2}&\cdots&g_{k,0},
\end{bmatrix},
\end{equation}
\begin{equation}\label{g-k-n}
\begin{split}
g_{k,0}=&\exp\left(-\pi\lambda_j\left(\frac{P_jB_j}{P_kB_k}\right)^{\frac{2}{\alpha_L}}\right.\\
&\left.\left({}_2F_1\left(\frac{2}{\alpha_L},M_L;1-\frac{2}{\alpha_L};-\frac{\theta M_{k,L}B_k}{G_0B_j}\right)-1\right)\right)
\end{split}
\end{equation}

Note that ${}_2F_1(\cdot)$ is the Gauss hypergeometric function \cite{integral-table},
\end{Corollary}
\emph{Proof:} See Appendix D.

\begin{Theorem}\label{theorem-moment}
Given that $u_0$ is associated with a LOS/NLOS BS in the $k$th tier, the $b$-th moment of the conditional success probability is given by
\begin{equation}\label{equation-moment}
\begin{split}
&\xi_{b,k,\rho}=\frac{1}{A_{k,\rho}}\int_{0}^{\infty}f_{L_{k,\rho}}(l_{k,\rho})\left(\sum_{\tau_1=0}^{b}\sum_{\tau_2=0}^{M_{\rho}\tau_1}\binom{b}{\tau_1}\binom{M_{\rho}\tau_1}{\tau_2}
\right.\\
&\left.(-1)^{\tau_1+\tau_2}e^{-s\sigma^2\zeta_{\rho}\tau_2}\prod_{j=1}^{K}\mathcal{L}_{I_{j,L}}(s\zeta_{\rho}\tau_2)\mathcal{L}_{I_{j,N}}(s\zeta_{\rho}\tau_2)\right)\text{d}l_{k,\rho}
\end{split}
\end{equation}
where $\zeta_{\rho}=(M_{\rho}!)^{-1/M_{\rho}}$, $\mathcal{L}_{I_{j,\nu}}(s\zeta_{\rho}\tau_2),\nu\in\{L,N\}$ is given by
\begin{equation}
\begin{split}
\mathcal{L}_{I_{j,\nu}}(s\zeta_{\rho}\tau_2)&=\exp\left(\int_{\frac{P_jB_j}{P_kB_k}l_{k,\rho}}^{\infty}\right.\\
&\left.\left(1-\left(1+\frac{\theta L_{k,\rho}M_{\rho}P_j\zeta_{\rho}\tau_2}{xP_kG_0M_{\nu}}\right)^{-M_{\nu}}\right)\Lambda_{j,\nu}(\text{d}x)\right)
\end{split}
\end{equation}

\end{Theorem}
\emph{Proof:} See Appendix E.

With the moments of the conditional success probability for each tier, the $b$-th moment of the conditional success probability for overall networks is given by
\begin{equation}
\xi_{b}=\sum_{k=1}^{K}\sum_{\rho\in\{L,N\}}A_{k,\rho}\xi_{b,k,\rho}
\end{equation}
By applying the Gil-Pelaez theorem, the meta distribution of the SIR for a CCU is given by
\begin{equation}\label{ccu-md}
\bar{F}_{\mathcal{P}}(y)=\frac{1}{2}+\frac{1}{\pi}\int_{0}^{\infty}\frac{\mathcal{J}\left(e^{-jt\log y}\xi_{jt}\right)}{t}\text{d}t,
\end{equation}
where $\mathcal{J}(z)$ is the imaginary part of $z$. Since the numerical evaluation of (\ref{ccu-md}) is cumbersome and it is difficult to obtain further insight, we utilize a beta distribution to approximate the meta distribution by matching the first and second moments, which can be easily obtained from the result in (\ref{equation-moment}):
\begin{equation}
\begin{split}
\xi_{1}=&\sum_{k=1}^{K}\sum_{\rho\in\{L,N\}}\int_{0}^{\infty}f_{L_{k,\rho}}(l_{k,\rho})\left(\sum_{\tau_2=1}^{M_{\rho}\tau_1}\binom{M_{\rho}}{\tau_2}
(-1)^{\tau_2+1}\right.\\
&\left.e^{-s\sigma^2\zeta_{\rho}\tau_2}\prod_{j=1}^{K}\mathcal{L}_{I_{j,L}}(s\zeta_{\rho}\tau_2)\mathcal{L}_{I_{j,N}}(s\zeta_{\rho}\tau_2)\right)\text{d}l_{k,\rho}
\end{split}
\end{equation}
\begin{equation}
\begin{split}
\xi_{2}&=\sum_{k=1}^{K}\sum_{\rho\in\{L,N\}}\int_{0}^{\infty}f_{L_{k,\rho}}(l_{k,\rho})\left(
2\sum_{\tau_2=1}^{M_{\rho}\tau_1}\binom{M_{\rho}}{\tau_2}(-1)^{\tau_2+1}\right.\\
&\left.e^{-s\sigma^2\zeta_{\rho}\tau_2}\prod_{j=1}^{K}\mathcal{L}_{I_{j,L}}(s\zeta_{\rho}\tau_2)\mathcal{L}_{I_{j,N}}(s\zeta_{\rho}\tau_2)
+\sum_{\tau_2=1}^{2M_{\rho}}\binom{2M_{\rho}}{\tau_2}\right.\\
&\left.(-1)^{\tau_2}e^{-s\sigma^2\zeta_{\rho}\tau_2}\prod_{j=1}^{K}\mathcal{L}_{I_{j,L}}(s\zeta_{\rho}\tau_2)\mathcal{L}_{I_{j,N}}(s\zeta_{\rho}\tau_2)
\right)\text{d}l_{k,\rho}
\end{split}
\end{equation}

By matching the variance and mean of the beta distribution, i.e., $\xi_{2}-\xi_{1}^2$ and $\xi_{1}$, the approximated meta distribution of the SIR can be given by
\begin{equation}\label{ccu-md-approximation}
\bar{F}_{\mathcal{P}_{k,\rho}}\approx 1-I_y\left(\frac{\xi_{1}\beta}{1-\xi_{1}},\beta\right),\ y\in[0,1],
\end{equation}
where
\begin{equation}
\beta=\frac{(\xi_{1}-\xi_{2})(1-\xi_{1})}{\xi_{2}-\xi_{1}^2}
\end{equation}
and $I_y(a,b)$ is the regularized incomplete beta function
\begin{equation}
I_y(a,b)\triangleq\frac{\int_{0}^{y}t^{a-1}(1-t)^{b-1}\text{d}t}{\text{B}(a,b)}.
\end{equation}

\subsection{Mean Local Delay}
In this subsection, we first derive the mean local delay, then obtain the closed-form expression of the variance of the mean local delay, i.e., the network jitter.

\begin{Theorem}\label{theorem-delay}
Given $u_0$ is associated with a LOS/NLOS BS in the $k$th tier, the mean local delay is given by
\begin{equation}
\begin{split}
&\xi_{-1,k,\rho}=\frac{1}{A_{k,\rho}}\int_{0}^{\infty}f_{L_{k,\rho}}(l_{k,\rho})\left(\sum_{\tau_1=0}^{\infty}\sum_{\tau_2=0}^{M_{\rho}\tau_1}\binom{M_{\rho}\tau_1}{\tau_2}
\right.\\
&\left.(-1)^{\tau_2}e^{-s\sigma^2\zeta_{\rho}\tau_2}\prod_{j=1}^{K}\mathcal{L}_{I_{j,L}}(s\zeta_{\rho}\tau_2)\mathcal{L}_{I_{j,N}}(s\zeta_{\rho}\tau_2)\right)\text{d}l_{k,\rho}
\end{split}
\end{equation}
\end{Theorem}
\emph{Proof:} See Appendix F.

Accordingly, we can obtain the expression of the $-2$-nd moment of the conditional success probability as follows
\begin{equation}
\begin{split}
&\xi_{-2,k,\rho}=\frac{1}{A_{k,\rho}}\int_{0}^{\infty}f_{L_{k,\rho}}(l_{k,\rho})\left(\sum_{\tau_1=0}^{\infty}\sum_{\tau_2=0}^{M_{\rho}\tau_1}\binom{M_{\rho}\tau_1}{\tau_2}
(-1)^{\tau_2}\right.\\
&\left.(\tau_1+1)e^{-s\sigma^2\zeta_{\rho}\tau_2}\prod_{j=1}^{K}\mathcal{L}_{I_{j,L}}(s\zeta_{\rho}\tau_2)\mathcal{L}_{I_{j,N}}(s\zeta_{\rho}\tau_2)\right)\text{d}l_{k,\rho}.
\end{split}
\end{equation}

With the -1-st nd -2-nd of the conditional success probability given $u_0$ is associated with a LOS/NLOS BS in each tier, the mean local delay and the -2-nd moment of the conditional success probability are given by
\begin{equation}
\xi_{-1}=\sum_{k=1}^{K}\sum_{\rho\in\{L,N\}}A_{k,\rho}\xi_{-1,k,\rho}
\end{equation}
\begin{equation}
\xi_{-2}=\sum_{k=1}^{K}\sum_{\rho\in\{L,N\}}A_{k,\rho}\xi_{-2,k,\rho}
\end{equation}

Therefore, the network jitter can be obtained by computing $\xi_{-2}-\xi_{-1}^2$.
\subsection{Moments of Conditional Rate Coverage}
The data rate depends on the total number of users served by a BS simultaneously. The probability mass function (PMF) of the number is given by
\begin{equation}
U_{k,\Upsilon}\triangleq\frac{1}{\Gamma(\upsilon)}\left(\frac{\lambda_ux}{3.5\lambda_k}\right)^{\upsilon-1}\frac{\Gamma(3.5+\upsilon)}{\Gamma(4.5)}
\left(1+\frac{\lambda_ux}{3.5\lambda_k}\right)^{-3.5-\upsilon}
\end{equation}
\begin{Theorem}
Given that $u_0$ is associated with a LOS/NLOS BS in the $k$th tier, the $b$-th moment of the conditional rate coverage probability can be expressed as
\begin{equation}
\xi_{R,b,k,\rho}\approx\sum_{\upsilon=1}^{\infty}U_{k,\Upsilon}\xi_{b,k,\rho}\left(2^{\frac{\epsilon\upsilon}{W}}-1\right)
\end{equation}
\end{Theorem}
\emph{Proof:} Given that $u_0$ is associated with a BS in the $k$th tier, the conditional rate coverage probability is given by
\begin{equation}
\mathcal{P}(\epsilon)=\mathbbm{P}(R_k>\tau|\Phi)
\end{equation}
Therefore, the $b$-th moment of the conditional rate coverage probability in the $k$th tier is given by
\begin{equation}
\begin{split}
\xi_{R,b,k,\rho}(\epsilon)&=\mathbbm{E}\left[\mathbbm{P}\left(R_k>\epsilon|\Phi\right)^{b}\right]\\
&=\mathbbm{E}_{\Upsilon_k}\left[\mathbbm{E}\left[\mathbbm{P}\left(R_k>\epsilon|\Phi,\Upsilon_k\right)^b\right]\right]\\
&=\sum_{\upsilon=1}^{\infty}U_{k,\Upsilon}\mathbbm{E}\left[\mathbbm{P}\left(\text{SINR}>2^{\frac{\epsilon\upsilon}{W}}-1|\Phi\right)^b\right]\\
&=\sum_{\upsilon=1}^{\infty}U_{k,\Upsilon}\xi_{b,k,\rho}\left(2^{\frac{\epsilon\upsilon}{W}}-1\right)
\end{split}
\end{equation}

\section{Simulation Results}
In this section, we consider a two-tier mmWave heterogeneous network, where the pico base station (PBS, Tier $k$=2) is overlaid with the micro base station (MBS, Tier $k$=1). The impact of the physical layer parameters on the association probability is provided. Then the impact of key network parameters on the mean and variance of the STP as well as the meta distribution are presented. Unless otherwise stated, the parameters are set as listed in the following table.

\newcommand{\tabincell}[2]{\begin{tabular}{@{}#1@{}}#2\end{tabular}}  
\begin{table}[htbp]
\centering
\caption{\label{tab:test}System Parameters}
\begin{tabular}{|c|c|c|}
\hline
$\textbf{Parameters}$&$\textbf{Values}$\\
\hline
Transmit power & \tabincell{c}{$P_1=43$dBm, $P_2=33$dBm} \\
\hline
Bias factor & $B_1=1$, $B_2=1$\\
\hline
Path loss exponent & $\alpha_{L}=2$, $\alpha_{N}=4$ \\
\hline
Density & \tabincell{c}{$\lambda_1=5/(500^2\pi)$, $\lambda_2=10/(500^2\pi)$}\\
\hline
Blockage parameter & \tabincell{c}{$\beta_1=0.006$, $\beta_2=0.024$}\\
\hline
Nakagami parameter& $ \rho_{L}=3$, $\rho_{N}=2$\\
\hline
Bandwidth & \tabincell{c}{$W=1$G}\\
\hline
Carrier frequency & 28GHz\\
\hline
$\kappa_{L}=\kappa_{N}$&$(F_c/4\pi)^2$\\
\hline
Zipf exponent & $\delta=0.6$\\
\hline
Caching capacity & \tabincell{c}{$F=50$, $C_1=35$,\\ $C_2=25$, $C_3=15$}\\
\hline
\end{tabular}
\end{table}

\begin{figure}
  \centering
  \includegraphics[width=3.5in]{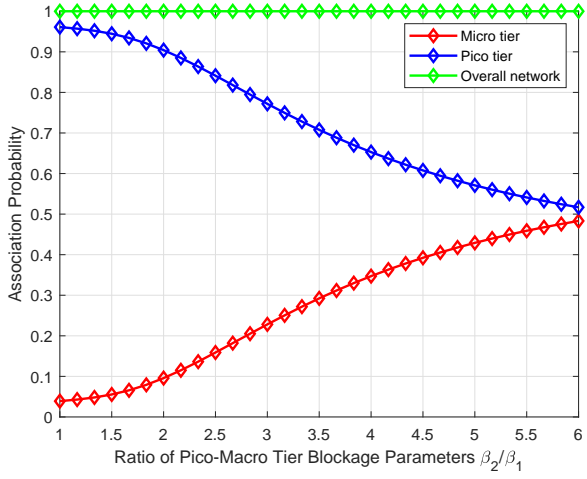}
  \caption{Association Probability versus ratio of micro-pico tier blockage parameters $\beta_2/\beta_1$.}\label{ap}
\end{figure}

Fig. \ref{ap} plots the association probability as the function of the blockage parameter. It can be observed that the association probability decreases with the blockage parameter of the associated tier. The explanation can be stated as follows. When the blockage parameter increases, the probability that the serving BS is LOS decreases and signal power received from the serving BS may be decreased, thereby leading to the reduction of the association probability.

\begin{figure}

\subfigure[Success probability versus SINR threshold]{
  \includegraphics[width=3.5in]{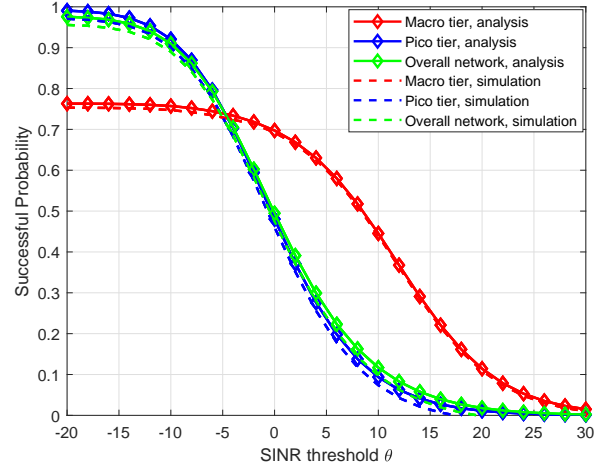}
  }
\subfigure[Variance of versus SINR threshold.]{
  \includegraphics[width=3.5in]{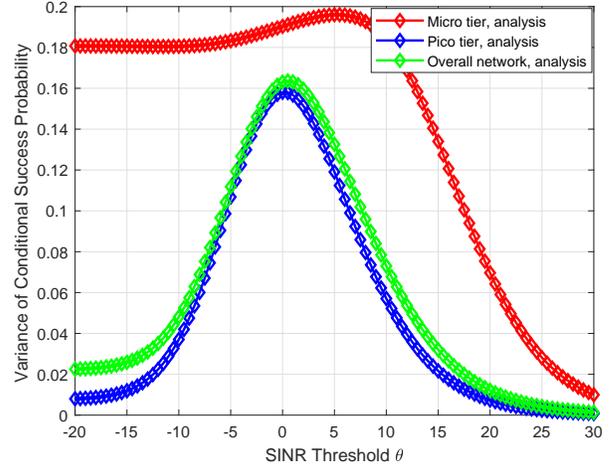}
  }
  \caption{The impact of SINR threshold $\theta$ on the per-tier success probability and the variance.}\label{sinr-threshold}
\end{figure}

Fig. \ref{sinr-threshold} plots the mean and variance of the success probability as the function of the SINR threshold $\theta$. The simulation results match the theoretical analysis well. The performance fluctuation can be reflected by the variance of the STP. A large variance corresponds to a large performance fluctuation and vice versa \cite{deng-SINR}. From Fig. \ref{sinr-threshold}(a), it can be observed that the STP decreases with the SINR. From Fig. \ref{sinr-threshold}(b), we can observe that there exists a maximum variance for both tiers. The $\theta$ corresponding to the maximum for the pico tier is almost identical to that of the overall network.

\begin{figure}

\subfigure[Success probability versus the ratio of pico-micro tier densities under different blockage parameters for the pico tier.]{
  \includegraphics[width=3.5in]{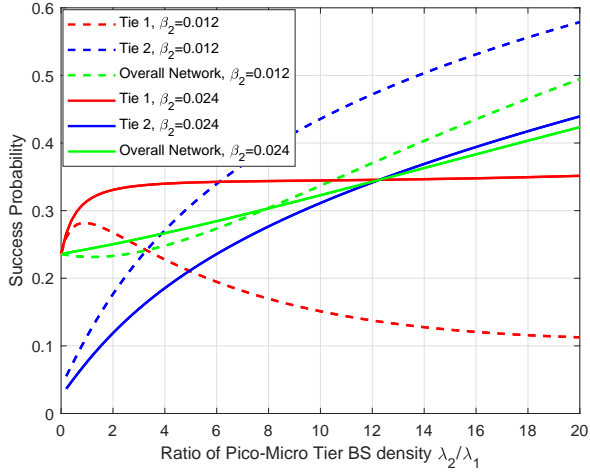}
  }
\subfigure[Variance versus the ratio of pico-micro tier densities under different blockage parameters for the pico tier.]{
  \includegraphics[width=3.5in]{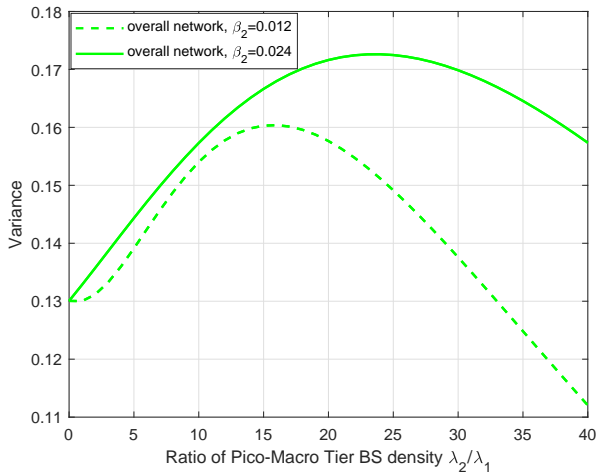}
  }
  \caption{The impact of the density of the pico tier $\lambda_2$ on the per-tier success probability and the variance.}\label{density}
\end{figure}

Fig. \ref{density} shows the impact of the density on the mean and variance of the success probability. From Fig. \ref{density}(a), it can be observed that when the density of pico tier increases, the success probability increases at start, then decrease gradually. The reason is that increasing the density of the pico tier motivates more users to be associated with the pico tier. Hence both the PBSs and MBSs have more per-user resources, resulting in a better performance. However, when the density of the pico tier further increases, the inter-tier interference experienced by $u_0$ increases and the performance of the micro tier degrades. It is also noteworthy that the success probability for the pico tier decreases with the blockage parameter of the pico tier $\beta_2$. Meanwhile, the success probability for the micro tier increases. This is because the probability that the serving BS is LOS decreases with $\beta_2$, causing a degradation of the inter-tier interference and the signal power experienced by $u_0$. We can observe from Fig. \ref{density}(b) that there exist a maximum variance of the success probability and the variance increases with $\beta_2$. In addition, the value of $\lambda_2/\lambda_1$ corresponding to the maximum increases with $\beta_2$.

\begin{figure}

\subfigure[Success Probability versus SINR threshold under different number of antenna elements]{
  \includegraphics[width=3.5in]{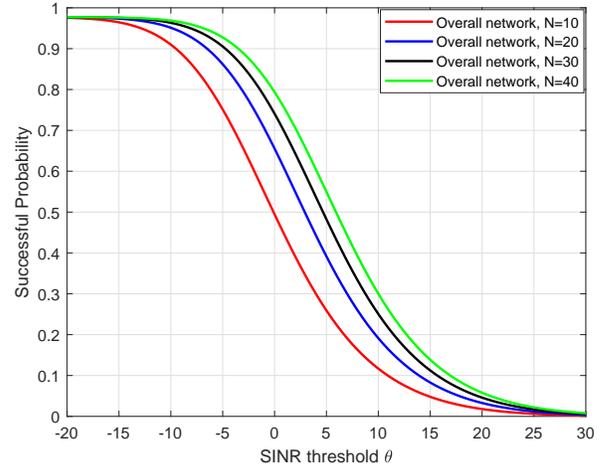}
  }
\subfigure[Variance versus SINR threshold under different number of antenna elements.]{
  \includegraphics[width=3.5in]{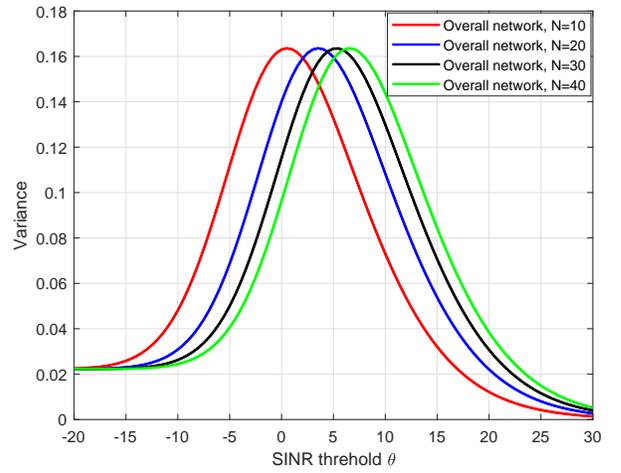}
  }
  \caption{The impact of the number of antenna elements $N$ on the per-tier success probability and the variance.}\label{antenna-number}
\end{figure}

Fig. \ref{antenna-number} plots the mean and variance of the success probability as the function of the number of antenna elements $N$. From Fig. \ref{antenna-number}(a), we can observe the success probability increases with $N$. The reason is that the signal power is enhanced with a larger number of antenna elements. From Fig. \ref{antenna-number}(b), we can observe there exists a maximum variance and the value of $\theta$ corresponding to the maximum increases with $N$. 

\begin{figure}
\subfigure[Success probability versus bias factor under different ratio of pico-micro tier densities]{
  \includegraphics[width=3.5in]{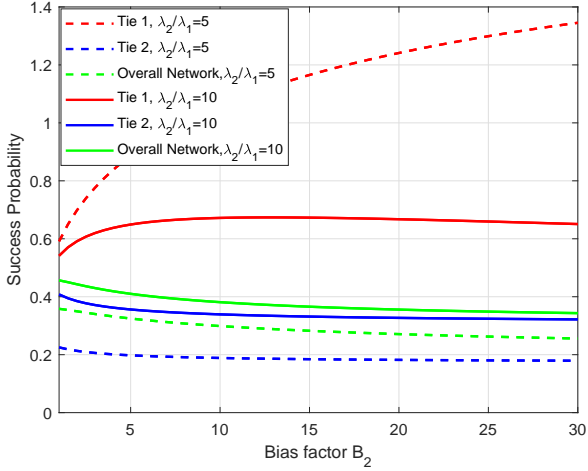}
  }
\subfigure[Variance versus bias factor under different ratio of pico-micro tier densities.]{
  \includegraphics[width=3.5in]{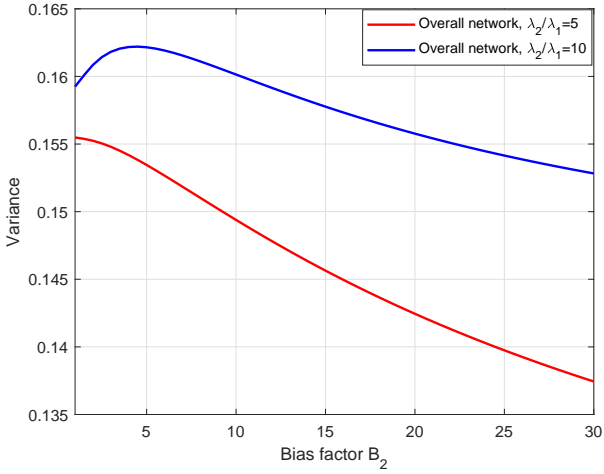}
  }
  \caption{The impact of bias factor the pico tier $B_2$ on the per-tier success probability and the variance.}\label{bias-factor}
\end{figure}

Fig.\ref{bias-factor} shows the impact of the bias factor of the pico tier $B_2$ on the mean and variance of the success probability under different ratios of pico-micro tier densities. From \ref{bias-factor}(a), it can be observed that success probability for the micro tier increases with $B_2$. Meanwhile, the success probability of the overall network decreases. This can be explained as follows. When the bias factor of the pico tier increases, more users are offloaded to the pico tier and the MBSs have more per-user resources, resulting in a better performance. However, with biasing, users are associated with the BSs not providing the strongest received signal power, leading to a slight decrease in the overall network performance. From \ref{bias-factor}(b), we observe that variance of the success probability decreases with $B_2$. In addition, the variance of success probability for the ratio of pico-micro tier densities $\lambda_2/\lambda_1=10$ is larger than that for $\lambda_2/\lambda_1=5$. This phenomenon indicates that compared with increasing the density of the PBSs, offloading the users to the pico tier by biasing can reduce the performance fluctuation while only causing a slight degradation of the overall network performance.

\begin{figure}
  \centering
  \includegraphics[width=3.5in]{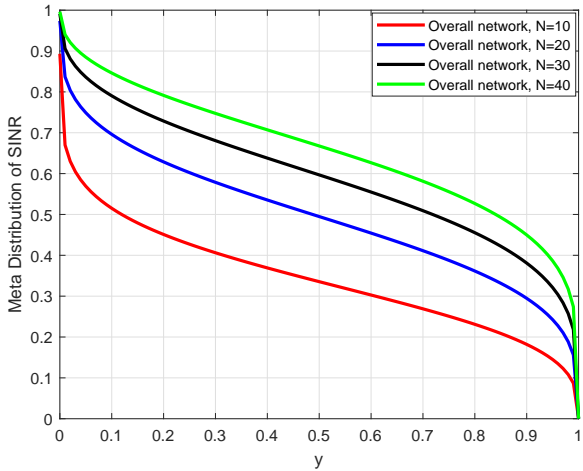}
  \caption{Meta distribution versus $y$ under different numbers of antenna elements $N$.}\label{md-antenna-number}
\end{figure}

\begin{figure}
  \centering
  \includegraphics[width=3.5in]{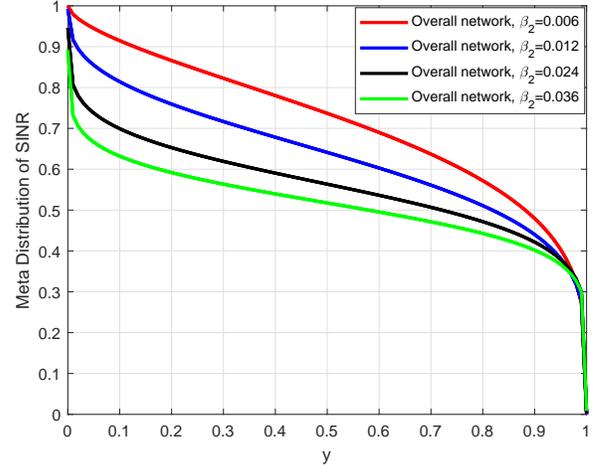}
  \caption{Meta distribution versus $y$ under different blockage parameters of the pico tier.}\label{md-blockage-parameter}
\end{figure}

Fig. \ref{md-antenna-number} analyzes the impact of the number of antenna element $N$ on meta distribution of the SINR. The results is consistent with that in Fig. \ref{antenna-number}. With larger $N$, the proportion of the users with higher success probability increases. Fig. \ref{md-blockage-parameter} plots the meta distribution of the SINR as the function of $y$ under different blockage parameters of the pico tier. We observe that the proportion of the users with higher success probability decreases with the blockage parameter of the pico tier $\beta_2$. We take $y=0.5$ for example. When $\beta_2=0.006$, more than 70$\%$ users achieve the success probability of at least 0.5. However, when $\beta_2$ increase to 0.036, 52$\%$ users can achieve the success probability of at least 0.5.

\begin{figure}
  \centering
  \includegraphics[width=3.5in]{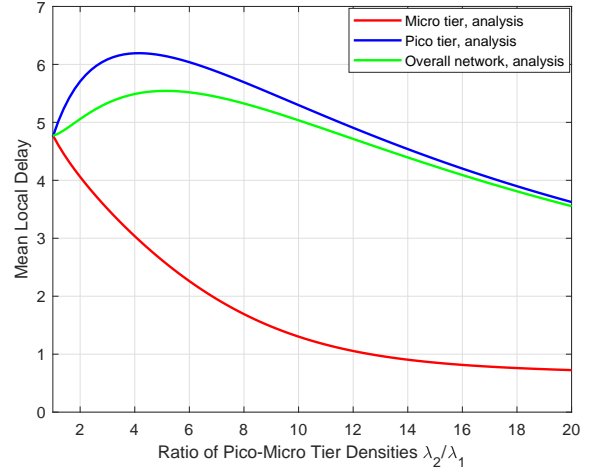}
  \caption{Mean local delay versus ratio of pico-micro tier densities.}\label{ld-density}
\end{figure}

Fig. \ref{ld-density} plots the mean local delay as the function of the ratio of the pico-micro tier densities $\lambda_2/\lambda_1$. It can be observed that there exists a maximum local delay for the pico tier and the overall network. In addition, the mean local delay for the micro tier decreases with $\lambda_2/\lambda_1$. The reason is that the more user are motivated to be associated with the pico tier when $\lambda_2/\lambda_1$ increases. Therefore, the MBSs have more per-user resources and the users are served more timely.

\section{Conclusion}
We focus on the meta distribution of the SINR in the mmWave heterogeneous networks. By utilizing stochastic geometry, we derive the moments of the conditional success probability, based on which the exact expression of meta distribution and its beta approximation are derived. In addition, key performance metrics, the success probability, the variance of the conditional success probability, the mean local delay and the network jitter are achieved.

\section{Appendix}
\subsection{Proof of Lemma \ref{lemma-intensity}}

The intensity measure for the BSs caching File $f$ in the $k$th tier under LOS probability function model can be computed as
\begin{equation}
\begin{split}
&\Lambda_k([0,x))
=\int_{0}^{\left(x/\kappa_L\right)^{1/\alpha_L}}2\pi\lambda_kve^{-\beta_kv}\,dv\\
&+\int_{0}^{\left(x/\kappa_N\right)^{1/\alpha_{N}}}2\pi\lambda_kv\left(1-e^{-\beta_kv}\right)\,dv\\
&=\pi\lambda_k(x/\kappa_N)^{2/\alpha_N}\\
&+2\pi\lambda_k\beta_k^{-2}\left(1-e^{-\beta_k(x/\kappa_L)^{1/\alpha_L}}
\left(1+\beta_k(x/\kappa_L)^{1/\alpha_L}\right)\right)\\
&-2\pi\lambda_k\beta_k^{-2}\left(1-e^{-\beta_k(x/\kappa_N)^{1/\alpha_N}}
\left(1+\beta_k(x/\kappa_N)^{1/\alpha_N}\right)\right).
\end{split}
\end{equation}

$\Lambda'([0,x))$ can be obtained by computing the derivative of $\Lambda([0,x))$ as follows
\begin{equation}
\begin{split}
&\Lambda'([0,x))
=2\pi\lambda_k\alpha_L^{-1}\kappa_L^{-2/\alpha_L}x^{2/\alpha_L-1}
e^{-\beta_k(x/\kappa_L)^{1/\alpha_L}}\\
&+2\pi\lambda_k\alpha_N^{-1}\kappa_N^{-2/\alpha_N}x^{2/\alpha_N-1}
\left(1-e^{-\beta_k(x/\kappa_N)^{1/\alpha_N}}\right)
\end{split}
\end{equation}

\subsection{Proof of Lemma \ref{lemma-association}}

The association probability is
\begin{equation}
\begin{split}
&A_{k,\rho}=\mathbbm{P}\left(P_kG_kL_{k,\rho}^{-1}\geq\max_{j,j\ne k}P_jG_jL_{j}^{-1}\right)\\
&\times\mathbbm{P}(L_{k,\bar{\rho}}>L_{k,\rho})\\
&=\int_{0}^{\infty}\mathbbm{P}\left(P_kG_kL_{k,\rho}^{-1}\geq\max_{j,j\ne k}P_jG_jL_{j}^{-1}\Big| l_{k,\rho}\right)\\
&\times\mathbbm{P}(L_{k,\bar{\rho}}>L_{k,\rho}| l_{k,\rho})f_{L_{k,\rho}}(l_{k,\rho})\text{d}l_{k,\rho}\\
&=\int_{0}^{\infty}\prod_{j=1,j\ne k}^{K}\mathbbm{P}(P_kG_kL_{k,\rho}^{-1}\geq P_jG_jL_{j,\rho}^{-1}\Big| l_{k,\rho})\\
&\mathbbm{P}(L_{k,\bar{\rho}}>L_{k,\rho}| l_{k,\rho})f_{L_{k,\rho}}(l_{k,\rho})\text{d}l_{k,\rho}\\
&=\int_{0}^{\infty}e^{-\sum_{j=1,j\ne k}^{K}\Lambda_{j}\left(\left[0,\frac{P_jG_j}{P_kG_k}l_{k,\rho}\right)\right)}
e^{-\Lambda_{k,\bar{\rho}}([0,l_{k,\rho}))}\\
&\times\Lambda_{k,\rho}^{'}([0,l_{k,\rho}))e^{-\Lambda_{k,\rho}([0,l_{k,\rho}))}\,dl_{k,\rho}\\
&=\int_{0}^{\infty}\Lambda_{k,\rho}^{'}([0,l_{k,\rho}))e^{-\sum_{j=1}^{K}\Lambda_{j}\left(\left[0,
\frac{P_jG_j}{P_kG_k}l_{k,\rho}\right)\right)}\,dl_{k,\rho},
\end{split}
\end{equation}
where $\rho,\bar{\rho}\in\{\text{LOS},\text{NLOS}\}$ and $\rho\ne\bar{\rho}$.

\subsection{Proof of Theorem \ref{theorem-sp}}
Given that $u_0$ is associated with a LOS/NLOS BS in the $k$th tier, the success probability can be expressed as
\begin{equation}
\begin{split}
\mathcal{P}_{k,\rho}&=\mathbbm{P}\left(\frac{P_kG_0h_{k,0}L_k^{-1}(r)}{\sigma^2+I}>\theta\right)\\
&=\mathbbm{E}_{I,s}\left[\frac{\Gamma(M_{\rho},s(\sigma^2+I))}{\Gamma(M_{\rho})}\right]\\
&=\sum_{m=0}^{M_{\rho}-1}\mathbbm{E}_{I,s}\left[e^{-s(\sigma^2+I)}\frac{(s(\sigma^2+I))^m}{m!}\right]\\
&=\mathbbm{E}_s\left[\sum_{m=0}^{M_{\rho}-1}\frac{(-s)^m}{m!}\mathcal{L}^{(m)}(s)\right]
\end{split}
\end{equation}
where $\mathcal{L}(s)$ is the Laplace transform of the interference and the noise and the superscript $(m)$ denotes the $m$-th derivative of $\mathcal{L}(s)$. Due to the independence of $K$ tiers, $\mathcal{L}(s)$ can be expressed as follows
\begin{equation}
\mathcal{L}(s)=\exp(-s\sigma^2)\prod_{j=1}^{K}\mathcal{L}_{I_{j,L}}(s)\mathcal{L}_{I_{j,N}}(s)
\end{equation}

We first compute the Laplace transform of the LOS interfering BSs $\mathcal{L}_{j,L}(s)$ as follows
\begin{equation}
\begin{split}
&\mathcal{L}_{I_{j,L}}(s)=\mathbbm{E}\left[\exp\left(-s\left(\sum_{i\in\Phi_{j}\backslash x_0}P_jh_{j,i}L_{j,i}^{-1}\right)\right)\right]\\
&=\prod_{i\in\Phi_{j}\backslash x_0}\mathbbm{E}\left[\exp\left(-sP_jh_{j,i}L_{j,i}^{-1}\right)\right]\\
&=\prod_{i\in\Phi_{j}\backslash x_0}\frac{1}{\left(1+\frac{sP_j}{M_L}L_{j,i}^{-1}\right)^{M_L}}\\
&=\exp\left(-\int_{\frac{P_jB_j}{P_kB_k}l_{k,\rho}}^{\infty}
\left(1-\left(1+\frac{sP_j}{M_L}x^{-1}\right)^{-M_L}\right)\Lambda_{j,\rho}(\text{d}x)\right)
\end{split}
\end{equation}

Note that $\mathcal{L}_{j,N}(s)$ can be obtained following the similar steps. Let $\mathcal{L}(s)=\exp(\eta(s))$ and we have $\mathcal{L}^{(1)}(s)=\eta^{(1)}(s)\mathcal{L}(s)$. Therefore, $\mathcal{L}_{{m}}(s)$ can be derived recursively following the formula of Leibniz for the product of two functions, which is given by
\begin{equation}\label{Laplace-recursive}
\mathcal{L}^{(m)}(s)=\frac{\text{d}^{m-1}}{\text{d}s}\mathcal{L}^{(1)}(s)=\sum_{n=0}^{m-1}\binom{m-1}{n}\eta^{(m-n)}(s)\mathcal{L}^{(n)}(s),
\end{equation}
where the $n$-th derivative of the $\eta(s)$ is given by
\begin{equation}\label{nth-derivative}
\begin{split}
&\eta^{(n)}(s)=-\sigma^2[2-n]^{+}-\sum_{j=1}^{K}\sum_{\nu\in\{L,N\}}\int_{\frac{P_jB_j}{P_kB_k}l_{k,\rho}}^{\infty}\\
&(-1)^n(M_{\nu})_n(P_jx^{-\alpha})^n
\left(1+sP_jx^{-\alpha}\right)^{-M_{\nu}-n}\Lambda_{j,\nu}(\text{d}x)\\
&=-\sigma^2[2-n]^{+}-\sum_{j=1}^{K}\sum_{\nu\in\{L,N\}}(-1)^n(M_{\nu}-1)_nP_js^{1-n}\\
&\int_{0}^{\frac{B_kM_{\rho}}{B_jG_0}}\frac{t^{n-2}}{(1+t)^{M_{\nu}+n}}\Lambda_{j,\nu}(\text{d}t)\\
\end{split}
\end{equation}

Letting $x_m=\frac{1}{n!}(-s)^{n}\mathcal{L}^{(n)}(s)$, the success probability can be rewritten as
\begin{equation}\label{P-xn}
\mathcal{P}_{k,\rho}=\mathbbm{E}\left[\sum_{m=0}^{M_{\rho}-1}x_m\right].
\end{equation}

Substituting $x_m=\frac{1}{n!}(-s)^{n}\mathcal{L}^{(n)}(s)$ into (\ref{nth-derivative}), we have
\begin{equation}\label{xn-recursive}
x_m=\sum_{n=0}^{m-1}\frac{m-n}{m}\left(\frac{(-s)^{m-n}}{(m-n)!}\eta^{(m-n)}(s)\right)x_n,
\end{equation}

Letting $q_{k,n}\triangleq\frac{(-s)^n}{n!}\eta^{(n)}(s)$, the recursive relationship of $x_m$ in (\ref{xn-recursive}) can be rewritten as $x_m=\sum_{n=0}^{m-1}\frac{m-n}{m}q_{k,m-n}x_n$. In order to solve $x_m$, two power series are defined as follows
\begin{equation}\label{Qz-Xz}
Q(z)\triangleq\sum_{m=0}^{\infty}q_{k,m}z^n,\ \ X(z)\triangleq\sum_{m=0}^{\infty}x_mz^m
\end{equation}

It can be easily verified that $X^{(1)}(z)=Q^{(1)}(z)X(z)$. In addition, we have $X(0)=\exp(Q(0))$. Therefore, the solution of (\ref{Qz-Xz}) is
\begin{equation}\label{Xz}
X(z)=\exp(Q(z))
\end{equation}

Combining (\ref{P-xn}), (\ref{Qz-Xz}) and (\ref{Xz}), we have
\begin{equation}
\begin{split}
\mathcal{P}_{k,\rho}&=\mathbbm{E}\left[\sum_{m=0}^{M_{\rho}-1}x_m\right]=\mathbbm{E}\left[\sum_{m=0}^{M_{\rho}-1}\left.\frac{1}{m!}X^{(m)}(z)\right|_{z=0}\right]\\
&=\mathbbm{E}\left[\sum_{m=0}^{M_{\rho}-1}\left.\frac{1}{m!}\frac{\text{d}^m}{\text{d}z^m}e^{Q(z)}\right|_{z=0}\right]
\end{split}
\end{equation}

From \cite{norm-1} and \cite{norm-1-original}, the first $M_{\rho}$ coefficients of $\exp(Q(z))$ form the first column of $\exp(\mathbf{Q}_{M_{\rho}})$. Therefore, the success probability is given by
\begin{equation}
\mathcal{P}_{k,\rho}=\mathbbm{E}\left[\left\|\exp\left(\mathbf{Q}_{M_{\rho}}\right)\right\|_1\right].
\end{equation}

\subsection{Proof of Corollary \ref{corollary-sp-los}}
Given that $u_0$ is associated with a LOS/NLOS BS in the $k$th tier, the success probability can be obtained following the similar steps with the proof of Theorem \ref{theorem-sp}. The only difference lies in that the interference from the NLOS BSs and the noise can be neglected and $\mathcal{L}(s)$ can be expressed as follows
\begin{equation}
\mathcal{L}(s)=\prod_{j=1}^{K}\mathcal{L}_{I_{j,L}}(s)
\end{equation}

We first compute the Laplace transform of the LOS interfering BSs $\mathcal{L}_{j,L}(s)$ as follows
\begin{equation}
\begin{split}
&\mathcal{L}_{I_{j,L}}(s)=\exp\left(-\frac{2\pi\lambda_j}{\alpha_L\kappa^{\frac{2}{\alpha_L}}}\right.\\
&\left.\int_{\frac{P_jB_j}{P_kB_k}l_{k,L}}^{\infty}
\left(1-\left(1+sP_jx^{-1}\right)^{-M_{\rho}}\right)x^{\frac{2}{\alpha_L}-1}\text{d}x\right)\\
&=\exp\left(-\pi\lambda_j\left(\frac{P_jB_jL_{k,L}}{P_kB_k\kappa_L}\right)^{\frac{2}{\alpha_L}}\right.\\
&\left.\left({}_2F_1\left(\frac{2}{\alpha_L},M_L;1-\frac{2}{\alpha_L};-\frac{\theta M_{k,L}B_k}{G_0B_j}\right)-1\right)\right)
\end{split}
\end{equation}
where the last equation follows from \cite{integral-table}. Then $\mathcal{L}(s)$ can be expressed as
\begin{equation}\label{Laplace-los}
\begin{split}
&\mathcal{L}(s)\overset{(a)}{=}\exp\left(-\pi \left(\frac{L_{k,L}}{\kappa_L}\right)^{\frac{2}{\alpha_L}}\left(\sum_{j=1}^{K}T-V\right)\right)
\end{split}
\end{equation}
where (a) follows from $V=\sum_{j=1}^{N}\lambda_j\left(\frac{P_jB_j}{P_kB_k}\right)^{\frac{2}{\alpha_L}}$ and $T$ is given in (\ref{T}).

Letting $y_m=\frac{1}{n!}(-s)^{n}\mathcal{L}^{(n)}(s)$, the success probability can be rewritten as
\begin{equation}\label{P-xn}
\mathcal{P}_{k,\rho}=\mathbbm{E}\left[\sum_{m=0}^{M_{\rho}-1}y_m\right].
\end{equation}

Substituting $x_m=\frac{1}{n!}(-s)^{n}\mathcal{L}^{(n)}(s)$ into (\ref{nth-derivative}), we have
\begin{equation}\label{y-g}
y_m=\sum_{n=0}^{m-1}\frac{m-n}{m}g_{k,m-n}y_n.
\end{equation}
where $g_{k,n}$ is given in (\ref{g-k-n}). Let $g_{k,0}=\sum_{j=1}^{K}W$, $G(z)\triangleq\sum_{m=0}^{\infty}g_{k,m}z^m$ and $Y(z)\triangleq\sum_{m=0}^{\infty}y_mz^m$. From \ref{y-g}, we have
\begin{equation}\label{Y-G-equation}
Y^{(1)}(z)=\pi G^{(1)}(z)Y(z)
\end{equation}

The solution of (\ref{Y-G-equation}) is $Y(z)=C\exp\left(\pi\left(\frac{L_{k,L}}{\kappa_L}\right)^{\frac{2}{\alpha_L}}G(z)\right)$. Then it remains to calculate $C$. From $Y(0)=y_0=\mathcal{L}(s)$, which is given in (\ref{Laplace-los}) and $G(0)=g_{k,0}$, $C$ can be obtained and $Y(z)$ is given by
\begin{equation}
Y(z)=\exp\left(\pi\left(\frac{L_{k,L}}{\kappa_L}\right)^{\frac{2}{\alpha_L}}
\left(G(z)+V-2g_{k,0}\right)\right)
\end{equation}

Since the PDF of $L_{k,L}$ is given by
\begin{equation}\label{path-loss-pdf-los}
\begin{split}
f_{L_{k,L}}(x)=&2\pi\left(\frac{x}{\kappa_L}\right)^{\frac{1}{\alpha_L}}\left(\sum_{j=1}^{K}\lambda_j
\left(\frac{P_jB_j}{P_kB_k}\right)^{\frac{2}{\alpha_L}}\right)\\
&\exp\left(-\pi\left(\frac{x}{\kappa_L}\right)^{\frac{2}{\alpha_L}}
\left(\frac{P_jB_j}{P_kB_k}\right)^{\frac{2}{\alpha_L}}\right)
\end{split}
\end{equation}

Therefore, $\mathbbm{E}_{L_{k,L}}[Y(z)]$ can be obtained as
\begin{equation}
\mathbbm{E}_{L_{k,L}}[Y(z)]=\frac{V}{2g_{k,0}-G(z)}.
\end{equation}

Then we have
\begin{equation}
\mathbf{P}=V\sum_{m=0}^{M_L-1}\frac{1}{m!}\frac{\text{d}^m}{\text{d}z^m}\left.\left(\frac{1}{2g_{k,0}-G(z)}\right)\right|_{z=0}.
\end{equation}

\subsection{Proof of Theorem \ref{theorem-moment}}
Given that $u_0$ is associated with a LOS/NLOS BS in the $k$th tier, the $b$-th moment of the conditional success probability can be expressed as
\begin{equation}
\begin{split}
&\xi_{b,k,\rho}=\mathbbm{P}\left(\frac{P_kG_0h_{k,0}L_k^{-1}(x)}{\sigma^2+I}>\theta\right)^b\\
&=\mathbbm{E}_{I,s}\left[\left(\frac{\Gamma(M_{\rho},s(\sigma^2+I))}{\Gamma(M_{\rho})}\right)^b\right]\\
&\overset{(a)}{=}\mathbbm{E}_{I,s}\left[\left(1-\frac{\gamma(M_{\rho},s(\sigma^2+I))}{\Gamma(M_{\rho})}\right)^b\right]\\
&\overset{(b)}{\approx}\mathbbm{E}_{I,s}\left[\left(1-\left(1-e^{-s(\sigma^2+I)\zeta_{\rho}}\right)^{M_{\rho}}\right)^b\right]\\
&\overset{(c)}{=}\mathbbm{E}_{I,s}\left[\sum_{\tau_1=0}^{b}\binom{b}{\tau_1}\left(-\left(1-e^{-s(\sigma^2+I)\zeta_{\rho}}\right)^{M_{\rho}}\right)^{\tau_1}\right]\\
&\overset{(d)}{=}\mathbbm{E}_{I,s}\left[\sum_{\tau_1=0}^{b}\sum_{\tau_2=0}^{M_{\rho}\tau_1}\binom{b}{\tau_1}\binom{M_{\rho}\tau_1}{\tau_2}
(-1)^{\tau_1+\tau_2}\exp(-s\sigma^2\zeta_{\rho}\tau_2)\right.\\
&\left.\prod_{j=1}^{K}\mathcal{L}_{I_{j,L}}(s\zeta_{\rho}\tau_2)\mathcal{L}_{I_{j,N}}(s\zeta_{\rho}\tau_2)\right]
\end{split}
\end{equation}
where (a) follows from $\Gamma(s)=\gamma(s,x)+\Gamma(s,x)$, (b) is from that the CDF of a Gamma random variable can be tightly upper bounded by $\frac{\gamma\left(M_{\rho},s(I+\sigma^2)\right)}{\Gamma(M_{\rho})}<[1-e^{-s(I+\sigma^2)\zeta_{\rho}}]$, (c) and (d) follows from the binomial expansion theorem, the definition of the Laplace transform of the interference and the noise and $\mathcal{L}(s)=\exp(-s\sigma^2)\prod_{j=1}^{K}\mathcal{L}_{I_{j,L}}(s)\mathcal{L}_{I_{j,N}}(s)$.

Since the PDF of the $L_{k,\rho}$ is obtained in (\ref{path-loss-pdf}), the proof of Theorem \ref{theorem-moment} can be completed by averaging over $L_{k,\rho}$.

\subsection{Proof of Theorem \ref{theorem-delay}}
Given that $u_0$ is associated with a LOS/NLOS BS in the $k$th tier, the mean local delay can be derived as
\begin{equation}
\begin{split}
&\xi_{-1,k,\rho}=\mathbbm{P}\left(\frac{P_kG_0h_{k,0}L_k^{-1}(r)}{\sigma^2+I}>\theta\right)^{-1}\\
&=\mathbbm{E}_{I,s}\left[\left(1-\frac{\gamma(M_{\rho},s(\sigma^2+I))}{\Gamma(M_{\rho})}\right)^{-1}\right]\\
&\approx\mathbbm{E}_{I,s}\left[\left(1-\left(1-e^{-s(\sigma^2+I)\zeta_{\rho}}\right)^{M_{\rho}}\right)^{-1}\right]\\
&\overset{(a)}{=}\mathbbm{E}_{I,s}\left[\sum_{\tau_1=0}^{\infty}(-1)^{\tau_1}\left(-\left(1-e^{-s(\sigma^2+I)\zeta_{\rho}}\right)^{M_{\rho}}\right)^{\tau_1}\right]\\
&=\mathbbm{E}_{I,s}\left[\sum_{\tau_1=0}^{\infty}\sum_{\tau_2=0}^{M_{\rho}\tau_1}\binom{M_{\rho}\tau_1}{\tau_2}
(-1)^{\tau_2}e^{-s\sigma^2\zeta_{\rho}\tau_2}\right.\\
&\left.\prod_{j=1}^{K}\mathcal{L}_{I_{j,L}}(s\zeta_{\rho}\tau_2)\mathcal{L}_{I_{j,N}}(s\zeta_{\rho}\tau_2)\right]
\end{split}
\end{equation}
where (a) follows from the binomial theorem for a negative integer power $(x+y)^n=\sum_{\tau=0}^{\infty}(-1)^{\tau}\binom{-n+\tau-1}{\tau}y^{n-\tau}x^{\tau}$.

\end{document}